# Machine learning aided atomic structure identification of interfacial ionic hydrates from AFM images

Binze Tang[1,2†], Yizhi Song[1,2†], Mian Qin[2†], Ye Tian[1,2†], Zhen Wei Wu[3], Ying Jiang[1,2,4,5,6*], Duanyun Cao[7,8*], Limei Xu[1,2,4,6*]

[1]*International Center for Quantum Materials, Peking University, Beijing, 100871, China*

[2]*School of Physics, Peking University, Beijing, 100871, China*

[3] *Institute of Nonequilibrium Systems, School of Systems Science, Beijing Normal University, 100875 Beijing, China*

[4] *Collaborative Innovation Center of Quantum Matter, Beijing, 100871, China*

[5] *CAS Center for Excellence in Topological Quantum Computation, University of Chinese Academy of Sciences, 100049, Beijing, China*

[6] *Interdisciplinary Institute of Light-Element Quantum Materials and Research Center for Light-Element Advanced Materials, Peking University, Beijing 100871, China*

[7]*Beijing Key Laboratory of Environmental Science and Engineering, School of Materials Science and Engineering, Beijing Institute of Technology, Beijing, 100081, China*

[8]*Beijing Institute of Technology Chongqing Innovation Center, Chongqing, 401120, China*

[†]These authors have contributed equally to this work and share co-first authorship.

[*]Corresponding author. E-mail: yjiang@pku.edu.cn, dycao@bit.edu.cn, limei.xu@pku.edu.cn

**ABSTRACT**

Relevant to broad applied fields and natural processes, interfacial ionic hydrates have been widely studied by ultrahigh-resolution atomic force microscopy (AFM). However, the complex relationship between AFM signal and the investigated system makes it difficult to determine the atomic structure of such a complex system from AFM images alone. Using machine learning, we achieved precise identification of the atomic structures of interfacial water/ionic hydrates based on AFM images, including the position of each atom and the orientations of water molecules. Furthermore, it was found that structure prediction of ionic hydrates can be achieved cost-effectively by transfer learning using neural network (NN) trained with easily available interfacial water data. Thus, this work provides an efficient and economical methodology which not only opens up avenues to determine atomic structures of more complex systems from AFM images, but may also help to interpret other scientific studies involving sophisticated experimental results.

**Introduction**

Interfacial ionic hydrates are ubiquitous in nature and are closely related to a variety of essential issues in applied fields and natural processes, from electrocatalytic processes[1-4], seawater desalination[5-8], to biological ion channels[9, 10] and chemical reactions[11, 12]. The structural information of ionic hydrates at the atomic level is crucial for elucidating the various extraordinary physical and chemical properties of the ionic hydrate/solid interface. In recent years, qPlus-based noncontact AFM with CO-decorated tip[13], capable of directly imaging water nanoclusters with submolecular resolution[14-18], has emerged as the most promising candidate for characterizing interfacial water network structure and dynamics at the atomic level[19].

However, in sharp contrast to the oxygen skeleton, which can be clearly reflected, identifying hydrogen atoms from AFM images is a challenge because they are barely visible[20]. Therefore, accurate structure identification, by combining density functional theory (DFT)-based stability calculations and probe particle method-based AFM simulations[21, 22], typically requires many trial-and-error processes to exclude a large number of possible structural models with different OH orientations.

Recently, machine learning (ML) has been used in various scientific and technological studies[23-28], such as structure identification[29-32], structure discovery[29, 33, 34], and electrostatic discovery[35] related to microscopy imaging. Especially for AFM imaging, a convolution neural network (CNN) has been applied by Alldritt et al to resolve the configuration of organic molecules[36]. In their study, the feeding data for ML were the simulated AFM images generated based on DFT optimized structures from the pre-existing databases. However, such an approach is not suitable for interfacial ionic hydration systems, since there is no ready database including ionic hydrate structures, and it is costly and impractical to acquire such large amounts of structures based on DFT calculations. In addition, this ML method compresses the 3-dimensional (3D) structure into 2-dimensional (2D) convolution and has no skip connections in NN architecture, which may lead to the loss of structural information during the NN processing process of AFM images. Since the signals of hydrogen atoms are weaker than that of other atoms (such as oxygen), this further increases the difficulty of identifying H. Therefore, it requires a cost-effective way to obtain input data, as well as a good structure representation and an effective NN architecture to distinguish H from other atoms.

Generally, the generation of input data consists of two key steps, structure acquisition and AFM image simulation. For structure acquisition, it requires sufficient phase space sampling of the target system to effectively introduce structural physics-induced biases into NN[24]. Obviously, the more accurate and richer the input structure, the higher the prediction accuracy. The DFT-based method has high accuracy, but cannot sample enough structures for training due to the great demands for computational resources[37]. As an alternative, the classical molecular dynamic (MD) simulations can provide many plausible structures economically and quickly[38]. At the same time, its lack of calculated energy accuracy has a minor impact on structure prediction and is therefore negligible in structure acquisition for ML.

In terms of AFM simulations, the calculations of the potential surface distribution of the system used for AFM simulation must be accurate enough to agree well with the experimental results. Without the need for DFT calculations to obtain the system electrostatic potential, the classical point charge model incorporating Lennard-Jones (L-J) interactions[21, 22] can be used to cost-effectively perform AFM simulations of interfacial water. However, when ions are added, this point charge method no longer accurately provides the potential surface for the ionic hydrate system, so DFT calculations become crucial. Fortunately, transfer learning[39] can be used to transfer knowledge from interfacial water systems to the interfacial ionic hydrate systems because their hydrogen bond (H-bond) networks share similar characteristics. Such usage can nicely reduce the need for interfacial ionic hydrate data, making the training of NN for ionic hydrates more economical.

Therefore, taking $Na^+$ hydrates, one of the most abundant alkali metal ions in nature, as an example, we designed an ML method with transfer learning to economically determine the structure of interfacial ionic hydrates from the AFM images. A NN was first trained on a large number of AFM images of interfacial water structures simulated via classical MD, and then retrained via transfer learning on AFM images of $Na^+$ hydrates simulated based on DFT-computed electrostatic potentials. Strikingly, using only thousands of simulated data of $Na^+$ hydrates, the prediction accuracy of the retrained NN can reach 95% for both sodium and oxygen, and 85% even for hydrogen. From the designed structure representations predicted by NNs, the positions of each atom and the orientations of water molecules can be easily identified, which has not been achieved by any ML method to date. The accuracy and efficiency[40] of this prediction far exceed the trial-and-error process of human experts. Furthermore, this economical ML method is also a fairly general workflow for structure prediction from AFM images, and can be extended from interfacial ionic hydrate systems to other complex systems, such as the surface of bulk ice, and even to identify information from the results of other spectroscopic experiments[34, 41-43].

**Results**

As illustrated in Fig. 1, the workflow consists of preliminary training and retraining for structure prediction for interfacial water (Fig. 1a) and interfacial $Na^+$ hydrates (Fig. 1b), respectively. During preliminary training, the NN, which is a standard 3D-Unet model[44, 45] (see Method), is fed with sufficient interfacial water data to learn to resolve the structure information of H-bond networks from AFM images. Then, via transfer learning, the NN is retrained with a small amount of expensive interfacial $Na^+$ hydrates data to achieve efficient structure identification of $Na^+$ hydrates economically. In both cases, the implemented learning schemes include data preparation, data labeling, NN training and performance evaluation. Data preparation, including phase space exploration based on classical MD simulations, structure sampling, AFM images simulation and data processing, is the key to transfer learning and is described in more detail below. For the identification of hydrogen atoms, the designed structure representation (data label) and the corresponding evaluating criteria will also be stated later.

**Data Preparation for Preliminary Training**

A large number of simulated AFM images of well-sampled different structures from the interfacial water layer on the Au(111) substrate were used in the preliminary NN training to ensure the robustness of the NN. Considering the large density variation in different layers[46], the structure sampling in the sampled phase space was performed by sliding a detection window of $2.5 \times 2.5 \times 0.3 nm^3$ along the direction parallel and perpendicular to the substrate (Fig. 1a, the first panel). We note that the detection window is large enough to cover various interfacial water structures in the XY plane, and ensure that the AFM can only probe one layer along the Z direction (see SI section 4 for details). Up to 80,000 2D water structures of different temperatures and densities were selected for AFM simulations, of which 70,000 and 10,000 were randomly selected for training and for validation, respectively. For each structure, 10 images were randomly selected from a set of simulated AFM images with a tip-sample distance (defined in methods) of 1.14 nm to 1.34 nm (10 pm intervals), and stacked as the input data for NN training. Data augmentation including cutout, noise and pixel-shift was implemented before NN training to prevent overfitting (see SI section 5 for details).

**The NN Output and Performance Assessment for Structure Prediction**

An ideal output of structure prediction should include the chemical identity and positions of the atoms. Considering that NN cannot effectively predict individual coordinate values and specific atom types[36], we designed an intuitive representation to encode molecular structures, namely the advanced vdW (a-vdW) spheres. In this representation, atoms are represented by van der Waals radius ($\sigma$) and dispersion energy ($\epsilon$) of the L-J interaction, as well as the charge $q$. Thus, a colored atom map (Fig. 2, right two panels) can be obtained as the NN output (see SI section 6 for details), from which atomic identities and positions can be easily distinguished. We stress that this representation also contains some 3D structure information, because the atomic circles are stacked sequentially with the increasing height, so that the uppermost one of the overlapping atomic circles represents the highest atom.

The per-pixel mean square error (MSE) calculated between the reference image and the predicted image was used as the loss function of NN training to thoroughly evaluate the predictions for each structure. Moreover, considering that the goal of NN is object detection, we designed an algorithm to locate the positions of all atoms and further calculated the prediction accuracy of each atomic species for intuitively evaluating the NN performance (see SI section 16 for details). For each atomic species, the prediction accuracy was calculated based on all predictions for it in the whole dataset, rather than averaging its prediction accuracy of each structure. This is because the small number of atoms in a particular structure results in discontinuity of the prediction accuracy and further leads to inaccurate statistical results. The error bar for the prediction accuracy was calculated from the data at the top and bottom 20% of the MSE ranking to account for the uneven performance of the NN on different structure. In addition, in order to avoid overestimating the NN performance, the parameter set in the prediction accuracy calculation algorithm was adjusted and cross-checked by comparing the automatically and manually calculated prediction accuracy. The prediction accuracy of machine calculation with the final chosen parameter set is slightly lower than that of manual calculation, ensuring the reliability of the prediction accuracy of the machine calculation.

**Structure prediction of Interfacial Water based on Preliminary Training**

After preliminary training for 60 epochs using 70,000 data of interfacial water on Au(111) substrate, the loss converges to 0.0025. Structural predictions of high-density and low-density interfacial water by the preliminary trained NN are shown in Fig. 2a and b, indicating good predictions in both the atomic identity and position. Furthermore, the orientation of interfacial water molecules can also be well predicted by the different hydrogen colors in the defined structure representation. Hydrogens colored in light blue, green, and brown represent H-flat pose (O-H parallel to the surface), H-up pose ( O-H pointing obliquely upward toward the surface), and H-down pose (O-H pointing obliquely downward toward the surface), respectively. Based on 1000 validation data, the prediction accuracy of oxygen and hydrogen reaches as high as 96.2% and 89.9%, respectively. Apparently, this high-resolution interpretation capability of the NN outperforms human experts, who can barely predict all hydrogen positions based on AFM images alone.

To gain insight into the NN performance and further analyze its insufficiency, we analyzed the best and worst 10% predictions in order of the network loss. It was found that the loss in nearly all predictions (more than 99%) are caused by the light-colored dots or shadows on the graph (see Fig. S6) which correspond to low-confidence predictions, except for completely failed predictions (less than 1%). While these light-colored dots and shadows may hinder the accuracy improvements, the predicted structure is still a plausible atomic arrangement and thus can serve as a valuable reference.

To test the generalization ability of the NN preliminarily trained with interfacial water structures on Au(111), the structure prediction was carried out for interfacial water on Pt(111) which has a smaller lattice constant and is more hydrophilic than Au(111)[47]. The test structures include 1000 interfacial water structures at different temperatures and distances from the Pt (111) surface, and their processing method is the same as that of the structures on Au(111) surface. The structure prediction tested on Pt(111) performs pretty well on both dense and dilute interfacial water (Fig. 2c and d), with a slightly larger loss (0.0030) than that of Au(111) (0.0025). After testing, the prediction accuracies of oxygen atoms and hydrogen atoms are 94.8% and 86.9%, respectively, indicating the generalization ability of the NN and further laying the foundation for transfer learning.

In addition, further analyses of the structure predictions of Pt(111) found that the predicted hydrogen atom positions sometimes shift along the H-bonds compared to the reference (see Fig. S7, marked with arrows). Specifically, the hydrogen that should belong to water molecule A was predicted to belong to water molecule B, which is connected to A by hydrogen bonding. We speculate that, due to the weak hydrogen signals in the AFM images, the NN cannot locate hydrogen atoms based solely on image features like other atoms with strong signals, such as oxygen. Instead of detecting atom positions directly from AFM images, the NN uses structural inference to predict the positions of hydrogen atoms based on rules learned from the phase space of the training data. However, water molecules on hydrophobic Au(111) are more likely to be H-up than that on hydrophilic Pt(111), and thus the lack of the relevant structures in the preliminary training dataset caused the lower prediction accuracy of hydrogen on Pt(111). Indeed, the prediction mismatch of hydrogen (see Fig. S7) was almost completely reduced by retraining NN with the 5000 data of interfacial water structures on Pt(111). Thus, it can be found that data with physics-induced biases is essential and effective for NN to accurately predict one specific system.

**Structure prediction of $Na^+$ hydrates with Transfer Learning**

For the interfacial $Na^+$ hydrates system, we used transfer learning to efficiently train NN by loading the parameters of the preliminarily trained-NN. $Na^+ \cdot nH_2O$ (n=3,4) on NaCl(001) was chosen as the basic structure to explore the phase space of $Na^+$ hydrates system, whose structures were randomly located within the detection window. Unlike interfacial water, AFM simulations of the ionic hydrate structures were performed using electrostatic potentials obtained from DFT calculations. After training for 50 epochs with data based only on 1500 $Na^+$ hydrate structures, the loss converged to 0.0009 and the NN prediction performed very well (Fig. 3, the first and second row; Fig. S5). The predicted accuracies for sodium, oxygen and hydrogen calculated on 500 validation data are 96%, 95.4%, and 85%, respectively.

To test the transferability of the retrained NN via transfer learning, this NN was applied to predict the structure of more complex $Na^+$ hydrates with more than four water molecules. The predicted structure matches the actual structure remarkably well (Fig. 3, the third row; Fig. S5), and the prediction accuracies for sodium, oxygen and hydrogen atom based on the testing dataset of 1000 complex $Na^+$ hydrates, are 82.3%, 91.7%, and 72.9%, respectively. In addition, experimental AFM images of $Na^+ \cdot 4H_2O$ were used for structure prediction to test the effectiveness of the NN. The corresponding predicted structures agree very well with the DFT-optimized geometry (Fig. 3, the fourth row), and its simulated AFM images are in perfect agreement with the experimental result (see Fig. S8). We note that the light red dots in the upper left corner of the $Na^+ \cdot 4H_2O$ in the prediction can be discarded, as it is a low confidence prediction. Therefore, it can be determined that NNs can provide valuable structure predictions even for complex $Na^+$ hydrate AFM images and experimental AFM images. Moreover, following the training procedures, our NN gives a good prediction for experimental data of $K^+$ hydrates, which indicates the general applicability of our method (See Fig. S10, S11).

Generally speaking, the water molecules in $Na^+$ hydrate can be divided into those around $Na^+$ and those above $Na^+$ according to their positions relative to $Na^+$. To further understand the prediction performance of NN on different water molecules, the prediction accuracy of these two types of water was also analyzed (Fig. S9) and significant differences were found between them. For water molecules around $Na^+$, the prediction of oxygen and hydrogen is good. Conversely, atoms in water molecules above $Na^+$ are difficult to be predicted by NN. This poor prediction may be due to the fact that the actual height of such systems exceeds one layer and thus goes beyond the trained 2D representation, making NN parsing difficult.

**Validity Assessment of the Transfer Learning**

To evaluate the effectiveness of transfer learning, we used different training datasets (with different data volumes, $N =200, 500, 1000$ and $1500$) to train NN with or without transfer learning. The prediction accuracy for these different training segments was calculated using the same validation dataset containing 500 $Na^+$ hydrate structures, and the best 100 and worst 100 data sorted by loss were calculated as the upper and lower bounds of the error bar (Fig. 4). When $N \geq 500$, the NN trained with transfer learning has a higher prediction accuracy of all atomic species than that without transfer learning (Fig. 4a and b, blue line). The prediction accuracy of the hydrogen based on the directly trained NN without transfer learning is almost zero (Fig. 4c, orange line), and only small patches of light gray near oxygen are presented (Fig. S12). However, with transfer learning of the preliminarily trained NN, the high prediction accuracy of hydrogen can be achieved with only a few thousand data.

Notably, when N< 500, the directly trained NN shows better performance in $Na^+$ prediction than that with transfer learning. We speculate that this phenomenon is due to two aspects. One is that the signal for $Na^+$ in the AFM image is the strongest, which is easy for a blank NN to learn. The other is that the retraining of preliminarily trained NN via transfer learning requires sufficient $Na^+$ hydrate data to get rid of previous local minima. Furthermore, this drawback of transfer learning soon disappears as N increases to 500, while the prediction accuracy of transfer learning does not increase much when N > 1000, suggesting that only a few hundred data is required to train an accurate NN in this case. Therefore, it can be concluded that the NN preliminarily trained with the interfacial water data can be transferred to the ionic hydrate system through transfer learning, so that it can be realized to train the NN with only a small amount of data to obtain the ability of high-precision prediction for more complex systems.

**Discussion**

In this work, we presented an economic method to identify the atomic structure of interfacial ionic hydrates from AFM images using transfer learning which solves the huge demand for high-cost DFT calculation in the data preparation. Through empirical potential-based MD simulations and AFM simulations based on simple point charges and L-J interactions, we first sacrificed a little precision to generate a large number of AFM images of interfacial water, and trained the NN of a 3D-unet with these data to capture the atomic structure of the H-bond network. Next, by retraining with only a few thousand hydrate structure data through transfer learning, the NN can achieve accurate structure prediction of all atoms in hydrates, outperforming the ability of human experts. More importantly, after comparing the performance of NN training on the hydrate structure with or without transfer learning, we verified the feasibility and validity of the transfer learning method in reducing the demand for computational resources. In addition, this method is transferable to different systems, as demonstrated by the good predictions for both interfacial water on different substrates (Au and Pt surface) and complex $Na^+$ hydrates with different water molecule numbers.

Importantly, this work is a good start in high resolution atomic structure prediction based on AFM images. There is room for improvement, for instance, experimental errors such as noise distribution and tip drifting need to be handled more carefully to further improve the prediction accuracy. With the rapid application of physical laws and geometry symmetries informed NNs[24, 48] in general drug de novo design[49], biomolecular structure prediction[50, 51] and molecule surface potential prediction[52, 53], we believe that our approach can determine 3D structures (e.g. ice) by introducing those NN architectures. Finally, it is worth noting that, our proposed ML method with transfer learning can also be extended from efficient structure prediction for AFM imaging to a broad range of applications for interpreting other experimental measurements[54], such as STM[55], SEM[30] and TEM[56].

## METHODS

### Neural Network Structure

The NN for structure prediction from AFM images was a standard 3D U-Net[44] model, a CNN widely used for biological and clinical image segmentation[45]. It consists of an encoder part for feature extracting, a decoder part for converting input (AFM images) to output (visual representation), and skip connections between the encoder and decoder parts, effectively preventing the feature loss of H-atom during the encoder process (See details in SI section 1).

The loss function for this network is the mean squared error (MSE),

$$\text{MSE}(y, \tilde{y}) = \frac{1}{N}\sum_{i=1}^{n}(y_i - \tilde{y}_i)^2,$$

where $N$ is the number of images, $n$ is the number of pixels in each image, and $y$ and $\tilde{y}$ are the NN output and data labels, respectively. For the gradient descent optimization, the Adaptive Moment Estimation (Adam) optimizer was used[57].

### Molecular Dynamic Simulations

Vapor-liquid-solid system was constructed and the periodic boundary conditions were applied in all directions. Water molecules are represented by rigid SPC/E model[58]. Au/Pt atoms, fixed throughout the simulation, interact with water molecules only through Lennard-Jonnes interactions[59]. The Lorentz−Berteloth Combination Rules[60] is used. (More details are in SI section 2)

### Simulations of AFM Images

The AFM images were simulated using a molecular mechanics model based on methods described in refs [21] and [22]. We perform AFM simulations to model the CO-tip based on the probe-particle tip model with the following parameters, effective lateral stiffness k = 0.50 N/m, atomic radius $R_c$ = 1.661 Å, and Q = -0.05 e (e is the elementary charge). These parameters can effectively reproduce most of the important features of experimental AFM images. It is important to note that small changes in the simulation parameters for training data do not significantly change the predictions on the experimental data. The tip height is defined as the distance between the outmost metal atom of the tip and the average height of O atoms of water molecules. For the interfacial water on the metal substrate, the charge distribution is described as a point charge on each atom, while for $Na^+$ and $K^+$ hydrates, the electrostatic potentials used in the AFM simulations are obtained from DFT calculations, where the substrate has negligible influence on the AFM image and is ignored to reduce computational costs (see SI section 3 for details). The parameters of the Lennard-Jones pairwise potentials for all elements are listed in Table S3.

**AFM Experiments**

The experimental images of $Na^+$ hydrates were taken from ref.[14]. The experimental method for $K^+$ hydrated on the Au surface is described in detail here. The Au(111) single crystal was purchased from MaTeck. The Au(111) surface was cleaned by repeated $Ar^+$ ion sputtering at 1 keV and annealing at about 700 K for multiple cycles. The SAES alkali-metal dispensers were degassed before evaporation (current $I_K$ = 6.5 A, t = 2 min). The alkali metal atoms were deposited on the clean Au(111) surface at room temperature (deposition current $I_K$ = 6.3 A, t = 1 min). The ultrapure $H_2O$ (Sigma Aldrich, deuterium-depleted, 1 ppm) was used and further purified under vacuum by 3-5 freeze-and-pump cycles to remove remaining gas impurities. The water molecules were deposited on the Au(111) surface at 120 K. All the experiments were performed with a non-contact AFM system (Createc, Germany) at 5 K using a home-made qPlus sensor equipped with a tungsten (W) tip (spring constant $k_0 \approx 1,800$ N·m$^{-1}$, resonance frequency $f_0 \approx 28.7$ kHz, and quality factor $Q \approx 100,000$). All the AFM frequency shift ($\Delta f$) images were acquired with the CO-terminated tips in frequency modulation and constant-height mode. The oscillation amplitude of experimental AFM imaging is 100 pm.

**Data availability**

The data that support the findings of this study are available from the corresponding author upon reasonable request.

**Acknowledgement**

We thank the computational resources provided by the TianHe-1A supercomputer, the High Performance Computing Platform of Peking University, China, the Beijing Super Cloud Computing Center and the High-performance Computing Platform of Beijing Institute of Technology Chongqing Innovation Center. We thank Dr. Jinbo Peng for providing useful experimental images of $Na^+$ hydrates.

**Funding**

This work was supported by the the National Key R&D Program (No. 2021YFA1400501); the National Natural Science Foundation of China (No. 11935002 and 12204039); the National Postdoctoral Program for Innovative Talents (No. BX2021040) and the China Postdoctoral Science Foundation (No. 2021M690408); Beijing Institute of Technology Research Fund Program for Young Scholars (No. XSQD-202210007).

**Author Contributions**

B.T., Y. S., D. C., and L. X. designed the project, L. X. supervised the project. B.T., Y.S., M. Q, D.C., Z. W. carried out the simulations. Y.T. and Y.J. performed the AFM measurements. All authors contributed to analyze the data. B.T., Y.S., D.C. and L.X. wrote the manuscript and reply with the input of all other authors. The manuscript reflects the contributions of all authors.

*Conflict of interest statement.* None declared.

**Reference**

1. Suo LM, Borodin O, Gao T, et al.; "Water-in-salt" electrolyte enables high-voltage aqueous lithium-ion chemistries. *Science* 2015; **350**(6263): 938-943. doi: 10.1126/science.aab1595.

2. Resasco J, Chen LD, Clark E, et al.; Promoter Effects of Alkali Metal Cations on the Electrochemical Reduction of Carbon Dioxide. *Journal of the American Chemical Society* 2017; **139**(32): 11277-11287. doi: 10.1021/jacs.7b06765.

3. Liu TC, Lin LP, Bi XX, et al.; In situ quantification of interphasial chemistry in Li-ion battery. *Nature Nanotechnology* 2019; **14**(1): 50-6. doi: 10.1038/s41565-018-0284-y.

4. Huang BT, Rao RR, You SF, et al.; Cation- and pH-Dependent Hydrogen Evolution and Oxidation Reaction Kinetics. *Jacs Au* 2021; **1**(10): 1674-1687. doi: 10.1021/jacsau.1c00281.

5. Celebi K, Buchheim J, Wyss RM, et al.; Ultimate Permeation Across Atomically Thin Porous Graphene. *Science* 2014; **344**(6181): 289-292. doi: 10.1126/science.1249097.

6. Tunuguntla RH, Henley RY, Yao YC, et al.; Enhanced water permeability and tunable ion selectivity in subnanometer carbon nanotube porins. *Science* 2017; **357**(6353): 792-796. doi: 10.1126/science.aan2438.

7. Joshi RK, Carbone P, Wang FC, et al.; Precise and Ultrafast Molecular Sieving Through Graphene Oxide Membranes. *Science* 2014; **343**(6172): 752-754. doi: 10.1126/science.1245711.

8. Zhang HC, Hou J, Hu YX, et al.; Ultrafast selective transport of alkali metal ions in metal organic frameworks with subnanometer pores. *Science Advances* 2018; **4**(2): eaaq0066. doi: 10.1126/sciadv.aaq0066.

9. Doyle DA, Cabral JM, Pfuetzner RA, et al.; The structure of the potassium channel: Molecular basis of K+ conduction and selectivity. *Science* 1998; **280**(5360): 69-77. doi: 10.1126/science.280.5360.69.

10. Payandeh J, Scheuer T, Zheng N, et al.; The crystal structure of a voltage-gated sodium channel. *Nature* 2011; **475**(7356): 353-U104. doi: 10.1038/nature10238.

11. Cao DY, Song YZ, Peng JB, et al.; Advances in Atomic Force Microscopy: Weakly Perturbative Imaging of the Interfacial Water. *Frontiers in Chemistry* 2019; **7**: 626. doi: 10.3389/fchem.2019.00626.

12. Cao DY, Song YZ, Tang BZ, et al.; Advances in Atomic Force Microscopy: Imaging of Two- and Three-Dimensional Interfacial Water. *Frontiers in Chemistry* 2021; **9**: 745446. doi: 10.3389/fchem.2021.745446.

13. Bartels L, Meyer G, Rieder KH; Controlled vertical manipulation of single CO molecules with the scanning tunneling microscope: A route to chemical contrast. *Applied Physics Letters* 1997; **71**(2): 213-215. doi: 10.1063/1.119503.

14. Peng JB, Cao DY, He ZL, et al.; The effect of hydration number on the interfacial transport of sodium ions. *Nature* 2018; **557**(7707): 701-5. doi: 10.1038/s41586-018-0122-2.

15. Peng JB, Guo J, Hapala P, et al.; Weakly perturbative imaging of interfacial water with submolecular resolution by atomic force microscopy. *Nature Communications* 2018; **9**(122): 1-7. doi: 10.1038/s41467-017-02635-5.

16. Giessibl FJ; The qPlus sensor, a powerful core for the atomic force microscope. *Review of Scientific Instruments* 2019; **90**(1): 011101. doi: 10.1063/1.5052264.

17. Gross L, Mohn F, Moll N, et al.; The chemical structure of a molecule resolved by atomic force microscopy. *Science* 2009; **325**(5944): 1110-1114. doi: 10.1126/science.1176210.

18. Pavlíček N, Gross L; Generation, manipulation and characterization of molecules by atomic force microscopy. *Nature Reviews Chemistry* 2017; **1**(1): 1-11. doi: 10.1038/s41570-016-0005.

19. Ma RZ, Cao DY, Zhu CQ, et al.; Atomic imaging of the edge structure and growth of a two-dimensional hexagonal ice. *Nature* 2020; **577**(7788): 60-3. doi: 10.1038/s41586-019-1853-4.

20. Yurtsever A, Fernandez-Torre D, Gonzalez C, et al.; Understanding image contrast formation in TiO2 with force spectroscopy. *Physical Review B* 2012; **85**(12): 125416. doi: 10.1103/PhysRevB.85.125416.

21. Hapala P, Temirov R, Tautz FS, et al.; Origin of High-Resolution IETS-STM Images of Organic Molecules with Functionalized Tips. *Physical Review Letters* 2014; **113**(22): 226101. doi: 10.1103/PhysRevLett.113.226101.

22. Hapala P, Kichin G, Wagner C, et al.; Mechanism of high-resolution STM/AFM imaging with functionalized tips. *Physical Review B* 2014; **90**(8): 085421. doi: 10.1103/PhysRevB.90.085421.

23. Carleo G, Cirac I, Cranmer K, et al.; Machine learning and the physical sciences. *Reviews of Modern Physics* 2019; **91**(4): 045002. doi: 10.1103/RevModPhys.91.045002.

24. Karniadakis GE, Kevrekidis IG, Lu L, et al.; Physics-informed machine learning. *Nature Reviews Physics* 2021; **3**(6): 422-440. doi: 10.1038/s42254-021-00314-5.

25. Mater AC, Coote ML; Deep Learning in Chemistry. *Journal of Chemical Information and Modeling* 2019; **59**(6): 2545-2559. doi: 10.1021/acs.jcim.9b00266.

26. Li JL, Lim K, Yang HT, et al.; AI Applications through the Whole Life Cycle of Material Discovery. *Matter* 2020; **3**(2): 393-432. doi: 10.1016/j.matt.2020.06.011.

27. Kalinin SV, Ziatdinov M, Hinkle J, et al.; Automated and autonomous experiments in electron and scanning probe microscopy. *ACS nano* 2021; **15**(8): 12604-12627. doi: 10.1021/acsnano.1c02104.

28. Krull A, Hirsch P, Rother C, et al.; Artificial-intelligence-driven scanning probe microscopy. *Communications Physics* 2020; **3**(1): 1-8. doi: 10.1038/s42005-020-0317-3.

29. Sotres J, Boyd H, Gonzalez-Martinez JF; Enabling autonomous scanning probe microscopy imaging of single molecules with deep learning. *Nanoscale* 2021; **13**(20): 9193-U55. doi: 10.1039/D1NR01109J.

30. Ziatdinov M, Dyck O, Maksov A, et al.; Deep Learning of Atomically Resolved Scanning Transmission Electron Microscopy Images: Chemical Identification and Tracking Local Transformations. *Acs Nano* 2017; **11**(12): 12742-12752. doi: 10.1021/acsnano.7b07504.

31. Gordon O, D'Hondt P, Knijff L, et al.; Scanning tunneling state recognition with multi-class neural network ensembles. *Review of Scientific Instruments* 2019; **90**(10): 103704. doi: 10.1063/1.5099590.

32. Carracedo-Cosme J, Romero-Muñiz C, Pérez R; A Deep Learning Approach for Molecular Classification Based on AFM Images. *Nanomaterials* 2021; **11**(7): 1658. doi: 10.3390/nano11071658.

33. Zhang Y, Mesaros A, Fujita K, et al.; Machine learning in electronic-quantum-matter imaging

experiments. *Nature* 2019; **570**(7762): 484-90. doi: 10.1038/s41586-019-1319-8.

34. Gordon OM, Moriarty PJ; Machine learning at the (sub) atomic scale: next generation scanning probe microscopy. *Machine Learning: Science and Technology* 2020; **1**(2): 023001. doi: 10.1088/2632-2153/ab7d2f.

35. Oinonen N, Xu C, Alldritt B, et al.; Electrostatic discovery atomic force microscopy. *ACS nano* 2021; **16**(1): 89-97. doi: 10.1021/acsnano.1c06840.

36. Alldritt B, Hapala P, Oinonena N, et al.; Automated structure discovery in atomic force microscopy. *Science Advances* 2020; **6**(9): eaay6913. doi: 10.1126/sciadv.aay6913.

37. For organic molecules or crystals from existing databases, structural relaxation with DFT alone is relatively fast. However, the acquisition of more complex interfacial water structures requires exploration of the phase space of a large system (about 500 atoms with box size larger than 3×3×3 nm$^3$) for a sufficiently long time (>> 0.01 ns).

38. Gordon OM, Hodgkinson JEA, Farley SM, et al.; Automated Searching and Identification of Self-Organized Nanostructures. *Nano Letters* 2020; **20**(10): 7688-7693. doi: 10.1021/acs.nanolett.0c03213.

39. Zhuang F, Qi Z, Duan K, et al.; A comprehensive survey on transfer learning. *Proceedings of the IEEE* 2020; **109**(1): 43-76. doi: 10.48550/arXiv.1911.02685.

40. Based on 1000 AFM data test, the prediction of one AFM data takes about 0.3 seconds.

41. Wang C, Li H, Hao Z, et al.; Machine learning identification of impurities in the STM images. *Chinese Physics B* 2020; **29**(11): 116805. doi: 10.1088/1674-1056/abc0d5.

42. Azuri I, Rosenhek-Goldian I, Regev-Rudzki N, et al.; The role of convolutional neural networks in scanning probe microscopy: a review. *Beilstein journal of nanotechnology* 2021; **12**(1): 878-901. doi: 10.3762/bjnano.12.66.

43. Thomas JC, Rossi A, Smalley D, et al.; Autonomous scanning probe microscopy investigations over WS2 and Au {111}. *npj Computational Materials* 2022; **8**(1): 1-7. doi: 10.1038/s41524-022-00777-9.

44. Çiçek Ö, Abdulkadir A, Lienkamp SS, et al.; 3D U-Net: learning dense volumetric segmentation from sparse annotation. *International conference on medical image computing and computer-assisted intervention*: Springer, 2016, 424-432.

45. Litjens G, Kooi T, Bejnordi BE, et al.; A survey on deep learning in medical image analysis. *Medical image analysis* 2017; **42**: 60-88. doi: 10.1016/j.media.2017.07.005.

46. Buff F, Lovett R, Stillinger Jr F; Interfacial density profile for fluids in the critical region. *Physical Review Letters* 1965; **15**(15): 621. doi: 10.1103/PhysRevLett.15.621.

47. Ogasawara H, Brena B, Nordlund D, et al.; Structure and bonding of water on Pt (111). *Physical review letters* 2002; **89**(27): 276102. doi: 10.1103/PhysRevLett.89.276102.

48. Atz K, Grisoni F, Schneider G; Geometric deep learning on molecular representations. *Nature Machine Intelligence* 2021: 1-10. doi: 10.1038/s42256-021-00418-8.

49. Satorras VG, Hoogeboom E, Fuchs FB, et al.; E(n) Equivariant Normalizing Flows. *arXiv preprint arXiv:2105.09016* 2021. doi: 10.48550/arXiv.2105.09016.

50. Townshend RJ, Eismann S, Watkins AM, et al.; Geometric deep learning of RNA structure. *Science* 2021; **373**(6558): 1047-1051. doi: 10.1126/science.abe5650.

51. Jumper J, Evans R, Pritzel A, et al.; Highly accurate protein structure prediction with AlphaFold.

*Nature* 2021; **596**(7873): 583-589. doi: 10.1038/s41586-021-03819-2.

52. Behler J, Parrinello M; Generalized neural-network representation of high-dimensional potential-energy surfaces. *Physical review letters* 2007; **98**(14): 146401. doi: 10.1103/PhysRevLett.98.146401.

53. Zhang L, Han J, Wang H, et al.; Deep potential molecular dynamics: a scalable model with the accuracy of quantum mechanics. *Physical review letters* 2018; **120**(14): 143001. doi: 10.1103/PhysRevLett.120.143001.

54. Ge M, Su F, Zhao Z, et al.; Deep learning analysis on microscopic imaging in materials science. *Materials Today Nano* 2020; **11**: 100087. doi: 10.1016/j.mtnano.2020.100087.

55. Li J, Telychko M, Yin J, et al.; Machine vision automated chiral molecule detection and classification in molecular imaging. *Journal of the American Chemical Society* 2021; **143**(27): 10177-10188. doi: 10.1021/jacs.1c03091.

56. Rizvi A, Mulvey JT, Carpenter BP, et al.; A Close Look at Molecular Self-Assembly with the Transmission Electron Microscope. *Chemical Reviews* 2021; **121**(22): 14232-14280. doi: 10.1021/acs.chemrev.1c00189.

57. Kingma DP, Ba J; Adam: A method for stochastic optimization. *arXiv preprint arXiv:1412.6980* 2014. doi: 10.48550/arXiv.1412.6980.

58. Berendsen HJ, Postma Jv, van Gunsteren WF, et al.; Molecular dynamics with coupling to an external bath. *The Journal of chemical physics* 1984; **81**(8): 3684-3690. doi: 10.1063/1.448118.

59. Heinz H, Lin T-J, Kishore Mishra R, et al.; Thermodynamically consistent force fields for the assembly of inorganic, organic, and biological nanostructures: the INTERFACE force field. *Langmuir* 2013; **29**(6): 1754-1765. doi: 10.1021/la3038846.

60. Hansen JP, McDonald, I. R. *Theory of Simple Liquids*: Elsevier.

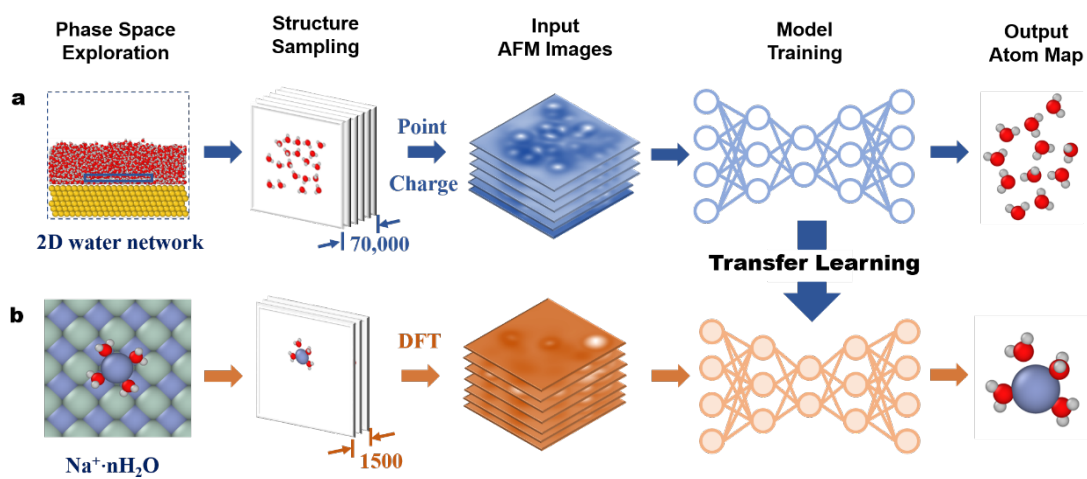

**Figure 1 | Schematic illustration of the overall framework of the training processes.** Training processes (a) and (b) are performed sequentially. First, 70,000 2D interfacial water network structures are sampled, based on which AFM images are simulated with point-charge electrostatic potentials as the input for the training process in (a). After training, simulated AFM images of 1,500 Na[+] hydrates structures are generated with DFT-calculated electrostatic potentials. These data are the input to the transfer learning process in (b). The output of the NN is a 2D representation of the molecular structure which looks similar to the atom map shown here.

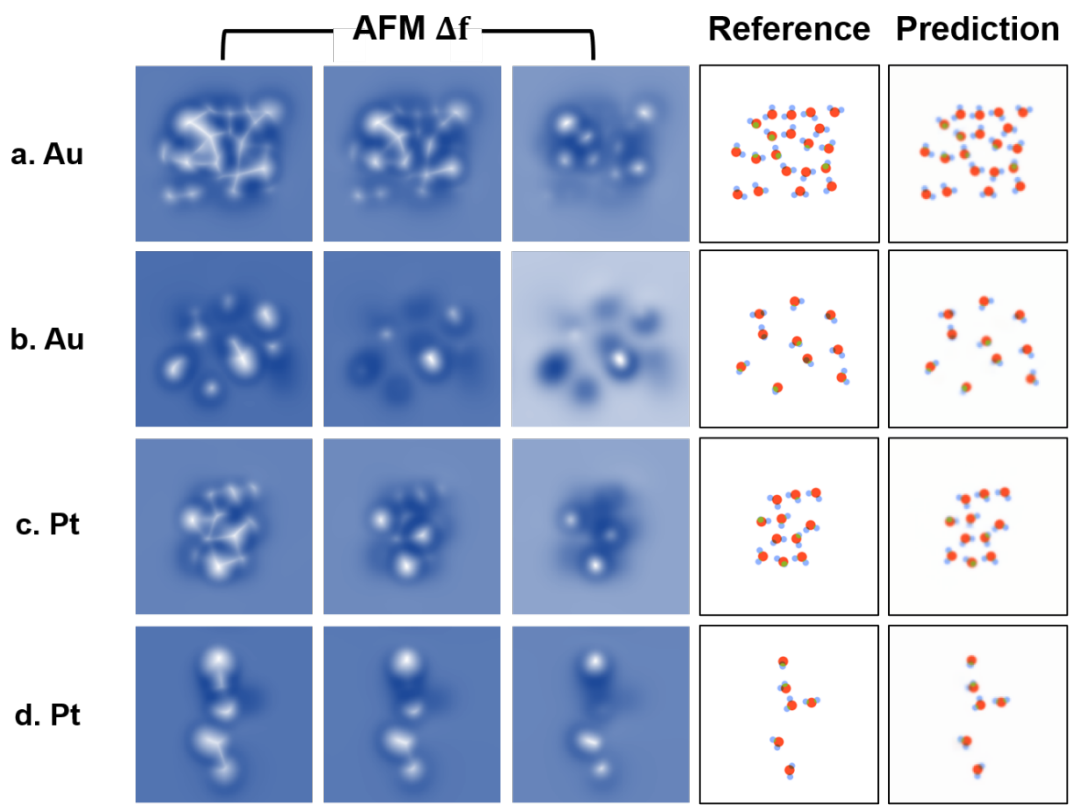

**Figure 2 | Examples of network prediction based on interfacial water simulation data.** (a) A dense water layer and (b) a dilute water layer on the Au surface selected from the validation dataset. (c) and (d) are the test data for dense and dilute water layers on the Pt surface to examine network transferability, respectively. Columns 1-3 are simulated AFM images (input data) with increasing tip-sample distances. Column 4 is the reference a-vdW spheres representation of the structure (label), and Column 5 is the prediction from the network. The red dots represent oxygen atoms, and the light blue, green, and brown dots represent hydrogen atoms at the same height as, above and below the oxygen atoms, respectively.

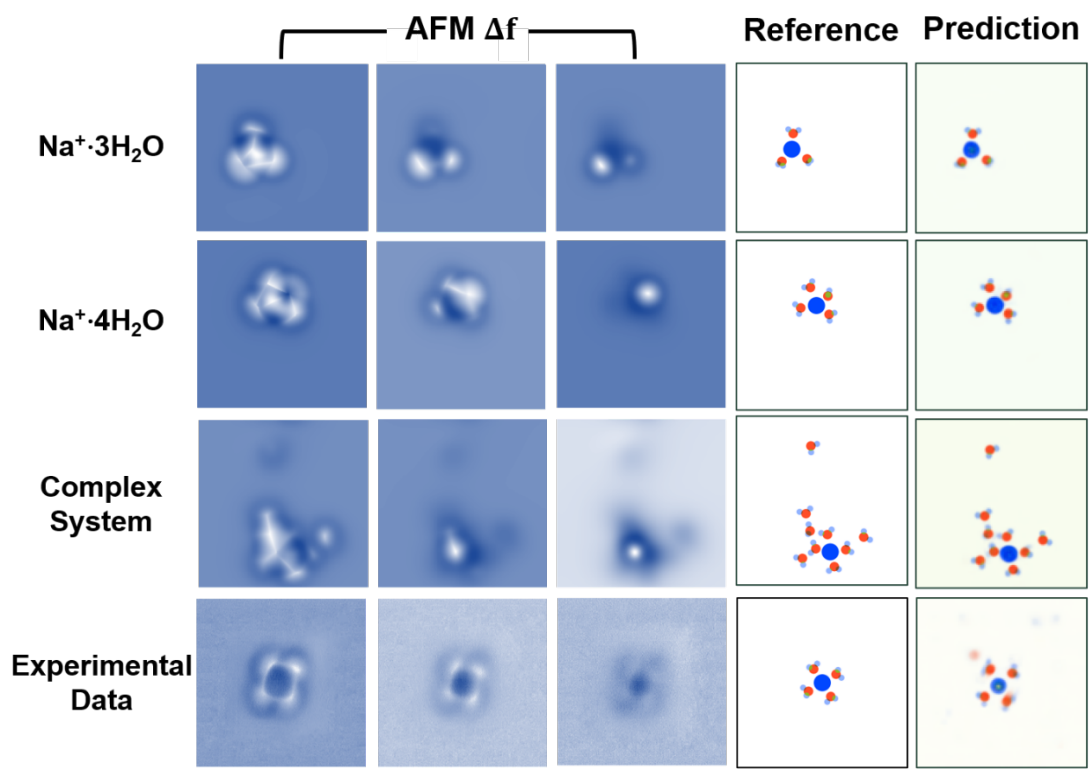

**Figure 3 | Examples of network prediction from simulated and experimental data for Na⁺ hydrates.** Rows 1 and 2 are the two basic structures, $Na^+ \cdot 3H_2O$ and $Na^+ \cdot 4H_2O$, selected from the validation dataset, respectively. Row 3 is the more complex sodium hydrate structure in the testing data to examine network transferability. Row 4 is the experimental data. Columns 1-3 are AFM images (input data) with increasing tip-sample distance. Column 4 is the reference a-vdW sphere representation of the structure (label), and column 5 is the prediction from the network. For the row of Experimental Data, the AFM images are obtained at different tip heights of 25 pm (left panel), 70 pm (middle panel) and 120 pm (right panel). The tip height of experimental AFM images is referenced to the STM set point on the NaCl surface (100 mV, 50 pA). Adapted with permission from Peng, et al. 2018. Nature 557 (7707): 701–705 (Peng, Cao, et al. 2018). The reference structure was validated by relaxing the predicted structure and comparing the AFM images simulated based on it with the experiment. The red and indigo spots represent oxygen and sodium atoms, respectively. The light blue, green, and brown spots represent hydrogen atoms at the same height as, above and below the oxygen atoms, respectively.

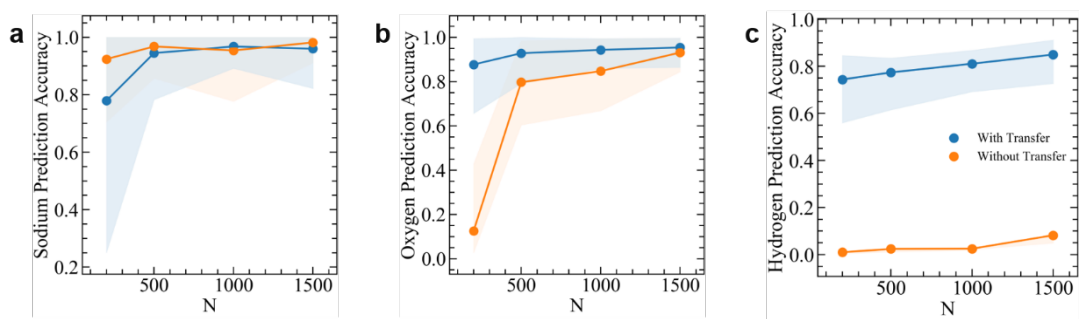

**Figure 4 | The prediction accuracy of a neural network as a function of the size of the training data.** (a), (b) and (c) are the positional accuracies of sodium, oxygen and hydrogen atoms, respectively. The blue and orange lines represent the training process with and without pretrained parameter loading, respectively. The solid line is calculated by counting the accuracy in all validation data, while the upper and lower bounds of the color patches are calculated from the data at the top and bottom 20% of the loss ranking, respectively.

# Supplementary Information: Machine Learning Aided Atomic Structure Identification of Interfacial Ionic Hydrates from AFM Images

Binze Tang[1,2,†], Yizhi Song[1,2,†], Mian Qin[2,†], Ye Tian[1,2,†], Zhen Wei Wu[3], Ying Jiang[1,2,4,5,6*], Duanyun Cao[7,8*], Limei Xu[1,2,4,6*]

[1]*International Center for Quantum Materials, Peking University, Beijing, 100871, China*

[2]*School of Physics, Peking University, Beijing, 100871, China*

[3] *Institute of Nonequilibrium Systems, School of Systems Science, Beijing Normal University, 100875 Beijing, China*

[4] *Collaborative Innovation Center of Quantum Matter, Beijing, 100871, China*

[5] *CAS Center for Excellence in Topological Quantum Computation, University of Chinese Academy of Sciences, 100049, Beijing, China*

[6] *Interdisciplinary Institute of Light-Element Quantum Materials and Research Center for Light-Element Advanced Materials, Peking University, Beijing 100871, China*

[7]*Beijing Key Laboratory of Environmental Science and Engineering, School of Materials Science and Engineering, Beijing Institute of Technology, Beijing, 100081, China*

[8]*Beijing Institute of Technology Chongqing Innovation Center, Chongqing, 401120, China*

[†]These authors have contributed equally to this work and share co-first authorship.

[*]Corresponding author. E-mail: yjiang@pku.edu.cn (Y. J.), dycao@bit.edu.cn (D. C.), limei.xu@pku.edu.cn (L.-X.)

**This PDF file includes:**

Section 1. Neural Network Structure.

Section 2. Calculation Details.

Section 3. Simulated AFM image of interfacial water with/without Au substrate

Section 4. Structure Sampling.

Section 5. Training Data Augmentation.

Section 6. Structure Representation.

Section 7. Training Process.

Section 8. NN Prediction of Interfacial Water.

Section 9. NN Prediction of hydration Sodium.

Section 10. Bad prediction of NN on interfacial water (Au surface) data.

Section 11. Correction of the hydrogen atom shift by transfer learning.

Section 12. Prediction for experimental AFM images of $Na^+ \cdot 4H_2O$.

Section 13. Prediction for Na hydrates whose water molecules are at different positions.

Section 14. Structure prediction of $K^+$ hydrates with Transfer Learning

Section 15. The prediction performance of NN with or without transfer learning.

Section 16. Object detection for each atom.

Figure S1. 3d U-Net architecture.

Figure S2. The simulated AFM image of interfacial water with/without Au substrate.

Figure S3. Structure representation.

Figure S4. Detailed prediction and reference representations of Fig. 2.

Figure S5. Detailed prediction and reference representations of Fig. 3.

Figure S6. Bad prediction of NN on interfacial water (Au surface) data.

Figure S7. Hydrogen atom shift in NN prediction of the water structure on Pt surface.

Figure S8. Validation of the predicted $Na^+ \cdot 4H_2O$ structure according to the experimental AFM images.

Figure S9. Prediction for Na hydrates whose water molecules are at different

positions.

Figure S10. Examples of network prediction of $K^+$ hydrates simulated and experimental data.

Figure S11. Validation of the predicted $K^+$ hydrates structure.

Figure S12. Prediction of Na hydrates by NN with or without transfer learning.

Figure S13. The prediction accuracy error bar in Fig. 4 with the manually collecting data.

Table S1. Force field parameters of SPC/E water, Au and Pt surface for interfacial water simulation.

Table S2.1. Force Field Parameters for Sodium Chloride.

Table S2.2. Force Field Parameters for Potassium Chloride

Table S2.3. Force Field Parameters for polarizable water model based on SPC/$\varepsilon$ model.

Table S3. Parameters of L-J pairwise potentials for AFM simulations.

Table S4. Parameters used in atom detection.

## 1. Neural Network Structure

The Neural Network (NN) for structure prediction from atomic force microscopy (AFM) images was a standard 3D U-Net[1] model, a CNN widely used for biological and clinical image segmentation[2]. It consists of an encoder part for feature extracting, a decoder part for converting input (AFM images) to output (visual representation), and skip connections between the encoder and decoder parts, effectively preventing the feature loss of H-atom during the encoder process (fig. S1). 10 of the gray scale AFM images at different tip-sample distances are stacked to a 3D image as the input to the network. The size of the input data is $1 \times 10 \times 128 \times 128$ where each number represents the channel, depth, height and width respectively. The output of the network is a 3 channels 2D image of size $3 \times 128 \times 128$ where each number represents the channel, height and width respectively. Each block in the encoder part contains two $3 \times 3 \times 3$ convolutional layers, each layer is preceded by a Group Normalization (GN) layer and followed by a leaky rectified linear unit (L-ReLU). And a maxpooling is added at the end of each block. The first two maxpoolings are $1 \times 2 \times 2$ in size to keep the AFM stack depth at 10, and the next two maxpoolings are $2 \times 2 \times 2$ in size to reduce the data depth to 2. In the decoder part, each block consists of an interpolation upsampling of the size of maxpooling in the corresponding block in the encoder, followed by two $3 \times 3 \times 3$ convolution layers, each of which is preceded by a GN layer and followed by a L-ReLU. Skip connections between layers with the same size in the encoder and decoder part provide the decoder with essential high-resolution features. After the decoder, there are two $3 \times 3 \times 3$ convolution layers and two $3 \times 3$ convolution layers. The data was flattened before the $3 \times 3$ convolution. Except for the last layer using ReLU, other convolution layers use L-ReLU. The slope of the L-ReLU is 0.001.

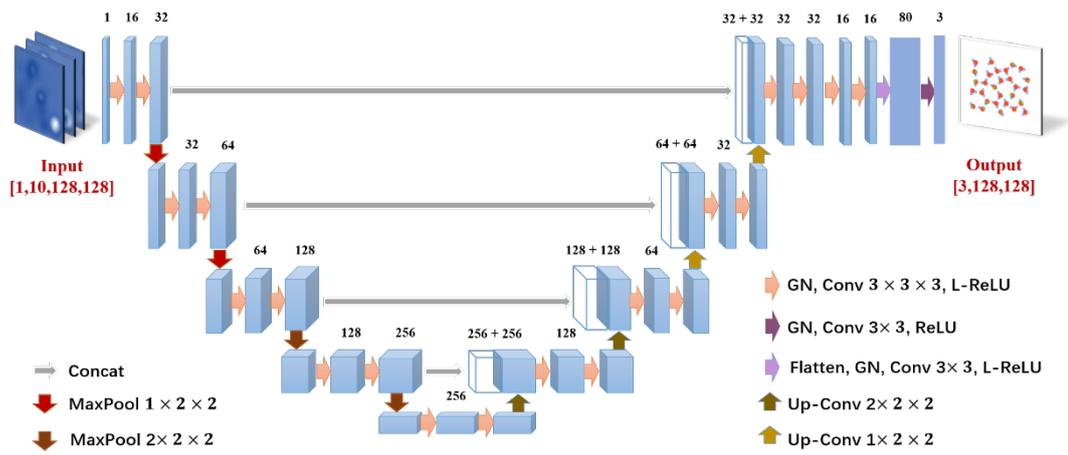

**Figure S1 | 3d U-Net architecture.** Schematic illustration of the model architecture. Blue cuboids and rectangles represent 3D and 2D feature maps, respectively. The number of channels is denoted above each feature map.

## 2. Calculation Details

2.1 The MD simulations of interfacial water.

The MD simulations of interfacial water are performed in the NVT ensemble using the LAMMPS software packages[3]. In the constructed system of the interfacial water on Au(111) surface, the size of the entire box is about $5.0 \times 5.0 \times 7.5 \ nm^3$, the Au(111) surface modeled by a six-layer slab is fixed at the bottom of the box, the water tank with a size of about $5.0 \times 5.0 \times 3.0 \ nm^3$ is placed right above the surface and the rest of the box is vacuum space. For the system of interfacial water on the Pt surface, we only replace the surface to Pt (111) and adjusted the box size according to the lattice difference between Pt and Au surface. The detailed force field parameters are listed in Table S1. MD simulations over a wide temperature range from 140K to 300K ($40K$ a step), were performed to sample various water phase spaces sufficiently. The simulation timestep is 1 fs and the total simulation time at a given $(N,V,T)$ is $5 \ ns$ where the first $2 \ ns$ is run for structure relaxation.

2.2 The MD simulations of ion hydrates.

The MD simulations of $Na^+$/$K^+$ hydrates were performed using the AMBER16 software package[4] with polarizable force fields[5][6] for $Na^+$ hydrates and the recently optimized force fields for $K^+$ hydrates[7]. The effectiveness of the force field for $Na^+$ hydrates has been demonstrated in our previous publication[8]. The force field parameters are shown in Table S2.1-2.3. A four-layered NaCl/KCl crystal ($18 \times 18 \times 4$ atomic number) with a (001) surface was used to support $Na^+$/$K^+$ hydrates. The MD simulation for each system was run no shorter than $1 \ ns$ for adequate configurations. The time step was set to 1 fs and the temperature is controlled using Langevin dynamics with a collision frequency of $0.1 ps^{-1}$. The bottom layer of the NaCl/KCl crystal was constrained by a force constant of $2000 \ kcal/(mol \cdot Å^2)$, and the periodic boundary conditions were applied in all directions. The SHAKE algorithm was used to constrain all bonds involving hydrogen atoms[9] and a cutoff of 1.0 nm was used for van der Waals interactions. A long-range dispersion correction based on an analytical integral, assuming an isotropic, uniform bulk particle distribution beyond the cut-off, was added to the van der Waals energy and pressure[4].

2.3 DFT calculation.

We calculate the electrostatic potential of $Na^+/K^+$ hydrate at Au (111) surface using DFT within the generalized gradient approximation of Perdew-Burke-Ernzerhof[10] (PBE-GGA), which has been shown to give good results for $H_2O$ hydrogen bonding[11] DFT calculations were performed using the Vienna ab initio simulation package (VASP)[12]. Projector augmented wave pseudopotentials were used with a cut-off energy of 550 eV for the expansion of the electronic wave functions. Van der Waals corrections for dispersion forces were considered by using the optB86b-vdW functional[13] [14].

**Table S1 Force field parameters of SPC/E water, Au and Pt surface for interfacial water simulation**

| Element | $q$ [e] | $\sigma$ [Å] | $\varepsilon$ [Kcal/mole] |
|---|---|---|---|
| H | +0.4238 | 0.0 | 0.0 |
| O | -0.8476 | 3.166 | 0.1553 |
| Au | 0.0 | 2.629 | 5.29 |
| Pt | 0.0 | 2.535 | 7.8 |

**Table S2.1 Force Field Parameters for Sodium Chloride**

| Model | $q$ [e] | $\lambda_c$ | $\sigma$ [Å] | $(\varepsilon/k_B)$ [K] |
|---|---|---|---|---|
| Na | +1 | 0.885 | 2.52 | 17.44 |
| Cl | -1 | 0.885 | 3.85 | 192.45 |

**Table S2.2 Force Field Parameters for Potassium Chloride**

| Model | $q$ [e] | $\sigma$ [Å] | $\varepsilon$[KJ/mol] |
|---|---|---|---|
| K | +1 | 2.791 | 2.223 |

| | | | | |
|---|---|---|---|---|
| Cl | -1 | | 5.029 | 0.02742 |

**Table S2.3 Force Field Parameters for polarizable water model based on SPC/ε model**

| model | $r_{OH}$ [Å] | $\Theta$ [deg] | $q_H$ ($q_O$) [e] | $\sigma$ [Å] | $(\varepsilon/k_B)$ [K] |
|---|---|---|---|---|---|
| SPC | 1 | 109.47 | 0.445 (-0.890) | 3.1785 | 84.9 |

**Table S3 Parameters of L-J pairwise potentials for AFM simulations**

| Element | $\varepsilon$ [meV] | r [Å] |
|---|---|---|
| H | 0.680 | 1.487 |
| O | 9.106 | 1.661 |
| Na | 10.0 | 1.40 |
| K | 10.0 | 1.50 |

## 3 Simulated AFM image of interfacial water with/without Au substrate

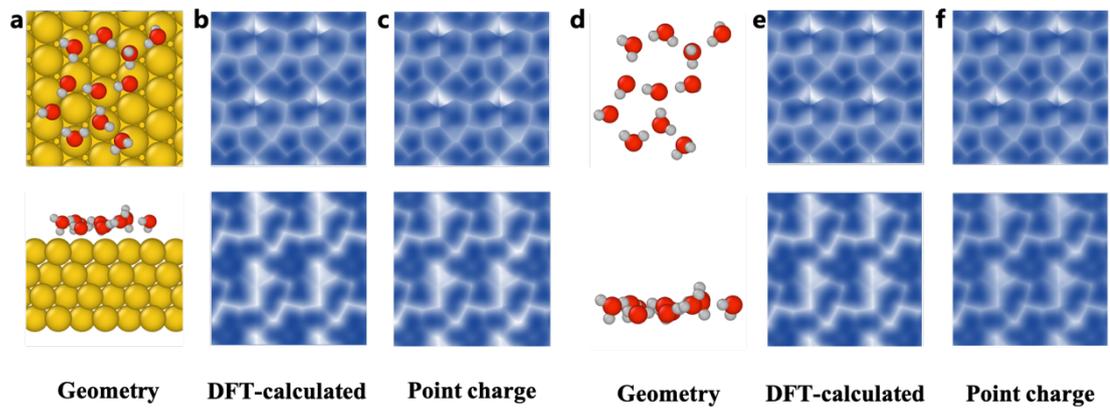

**Figure S2 | The simulated AFM image of interfacial water with/without Au substrate.** (a and d) The atomic structure of the interfacial water on Au substrate and the same structure without Au substrate, respectively (upper panel: top view; lower panel: side view). The red, white and golden spheres represent O, H and Au atoms, respectively. (b and c) The simulated AFM images of structure (a), where the electrostatic potential surface is obtained based on DFT (b) and point charge model (c), respectively. (e and f) The simulated AFM images of structure

(d), where the electrostatic potential surface is obtained based on DFT (e) and point charge model (f), respectively. It can be found that the effect of substrates on the AFM images was negligible, and the simulated AFM images with and without substrates are nearly identical.

## 4   Structure Sampling

We note that the detection window (Fig. 1a, the first panel) in size of $2.5 \times 2.5 \times 0.3 nm^3$ is large enough to cover the various interfacial water structures in the XY plane, and ensure that the AFM can only probe one layer along the Z direction. Thus, all of the structures are restricted in the window and their corresponding simulated AFM images have the same detection range in size $2.5 \times 2.5\ nm^2$.

For acquiring sufficient interfacial water structure, the detection window is initially settled at the first water layer and then slides in the direction parallel and perpendicular to the substrate. In XY plane, the number of steps of the detection window is 2 and the stride is $0.5\ nm$ for each step. While in Z direction, the number of steps and stride is 1 and $0.5\ nm$, respectively. To ensure the variety of the data, the water molecules within the distance $\delta b_x$ and $\delta b_y$ from the window boundary are neglected during each detection. The distance $\delta b_x$ and $\delta b_y$ are settled randomly, but can guarantee that at least one water molecule remains in the detection window. The configurations used for data acquisition is sampled from the MD simulations trajectories with a time interval of $10\ ps$.

## 5   Training Data Augmentation

To make the network more robust to experimental data and avoid overfitting, the input data is processed as follows.

a) Nosie. For each data, a random uniform noise or a gaussian noise is randomly selected and added to all the images.

A random uniform noise with an amplitude in a range of $[-\delta P, \delta P]$ is added to each image,

$$\delta P = c(P_{max} - P_{min}), \quad (S1)$$

where $P_{max}$ and $P_{min}$ is the maximum and minimum pixel value of an image,

and $c$ is the amplitude ratio.

A gaussian noise with an amplitude in a distribution $P_G$ is added to each image,

$$P_G(x) = 255 \times c' \frac{1}{\sqrt{2\pi}\sigma} \exp\left(-\frac{x^2}{2\sigma^2}\right), \quad (S2)$$

Where $c'$ is the amplitude ratio, $\sigma = 1$.

For training the interfacial water data, c=0.03, $c'$=0.01. When transfer learn the $Na^+/K^+$ hydrates for experimental prediction, c and $c'$ is randomly selected from a uniform distribution in range [0, 0.2] and [0, 0.15], respectively.

b) Pixel Shift. In experimental AFM images, for some technical reasons, the positions of specific structural features relative to the scanning boundary at different tip-sample distances could not remain exactly the same. Each image in our data is shifted a few pixels away in $\pm x$ and $\pm y$ directions. The movement distance is a random value of less than three pixels.

c) Cutout. Experimental AFM images inevitably have tip shifts, where there is something fuzzy. To simulate this, a rectangle with a gray value of the average of the entire image and a size less than 500 pixels is generated on each image to cover the image underneath.

d) Distance Randomization. 20 AFM images at 0.1 Å intervals are prepared. First, 15 consecutive images are selected. Then, 10 images are randomly selected from these to form a stack.

## 6 Structure Representation

We design a new structure representation named advanced vdW (a-vdW) spheres as the neural network output/ data label. As shown in fig. S3, it composed of three grayscale images representing atom type, positive charge and negative charge. They are the same size and are all the projection of the structure onto the $x - y$ plane. Since the interfacial structure is restricted in a box with a height of only 0.3 $nm$, its information in the $z$ direction is negligible, and the projection of all the atoms is sufficient to represent the structure to some extent. In our representation, each atom is described as a sphere and projected onto the $x-y$ plane in height order. If two atoms overlap, the higher atom cover the lower one. In atom-type images, the diameter $d$ and grayscale $g_1$ of atoms are related to their LJ potential parameters $\sigma$ and $\epsilon$. Here, in order to differentiate different elements for easier identification, $\epsilon$ of each element is scaled and transformed,

$$d = \sigma - k_1\left(1 - \left(\frac{\epsilon-\epsilon_{min}}{\epsilon_{max}}\right)^{\frac{1}{2}}\right), \quad (S3)$$

$$g_1 = 255 \times k_2\left(1 - \left(\frac{\epsilon-\epsilon_{min}}{\epsilon_{max}}\right)^{\frac{1}{2}}\right), \quad (S4)$$

where $\epsilon_{min}$ and $\epsilon_{max}$ are the minimum and maximum $\epsilon$ values among all elements, respectively, and $k_1 = 0.6$ and $k_2 = 0.7$ are the scale factors. All the LJ potential parameters are chosen from the OPLS Force Field[15]. For positive and negative charge images, the diameter of atom is defined as above, and the grayscale $g_2$ is scaled by charge,

$$g_2 = 255 \times \left(1 - \frac{|q|}{|q_{max}|}\right), \quad (S5)$$

where $q$ is the charge of the atom and $q_{max}$ is the charge of the most charged atom in the structure.

If we consider these three images as three channels of a color image and combine them, we get a colormap where each atom has a different color. According to the definition of this structure representation described above, the light blue, green, and brown hydrogens represent H-flat Pose (O-H parallel to the surface), H-up Pose ( O-H pointing obliquely upward to the surface), and H-down Pose (O-H pointing obliquely downward to the surface), respectively. This representation synthesizes the basic characteristics and position information of atoms. Similar to the structure atom map shown by a popular visualization software Ovito[16] (see fig. S3), this representation is highly readable and understandable.

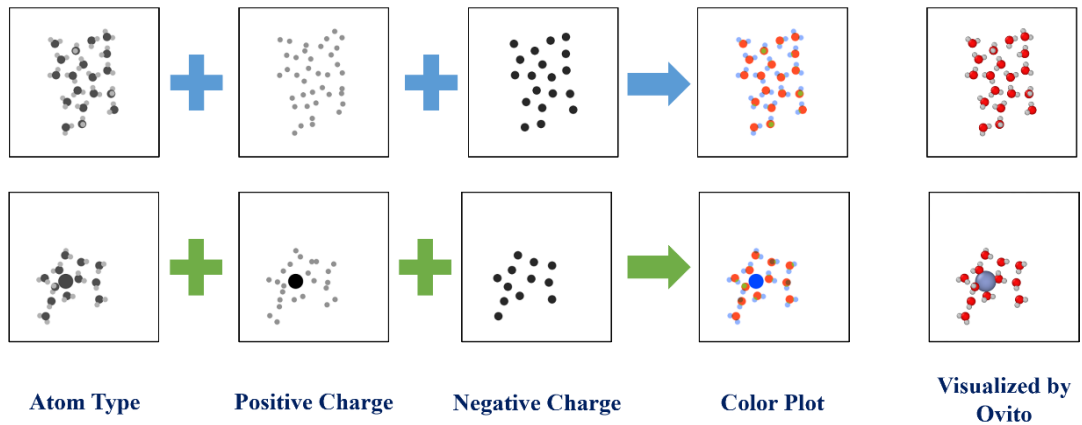

**Figure S3 | Structure representation**. The first three columns are grayscale representations of atom type, positive charge and negative charge, respectively. A colormap is the combination of three grayscale representations in RGB channels. The last column is a structure atom map visualized by the popular software Ovito for comparison. The first and second rows represent interfacial water and ion hydrate, respectively.

## 7 Training Process

The loss function is mean square error (MSE) and the optimizer for gradient descent is the Adaptive Moment Estimation (Adam)[17]. For the training of interfacial water data, the learning rate is 0.001 in the first 30 epochs, and then decreases to 0.0001 for the next 30 epochs. For the transfer learning of $Na^+$ hydrates, trained network parameters from the interfacial water dataset are used. The learning rate for the last two 2D $3 \times 3$ convolution layers and other layers are 0.001 and 0.0001, respectively. The loss converges after 60 epochs of training. The batch size we use throughout the training is 8.

## 8 NN Prediction of Interfacial Water

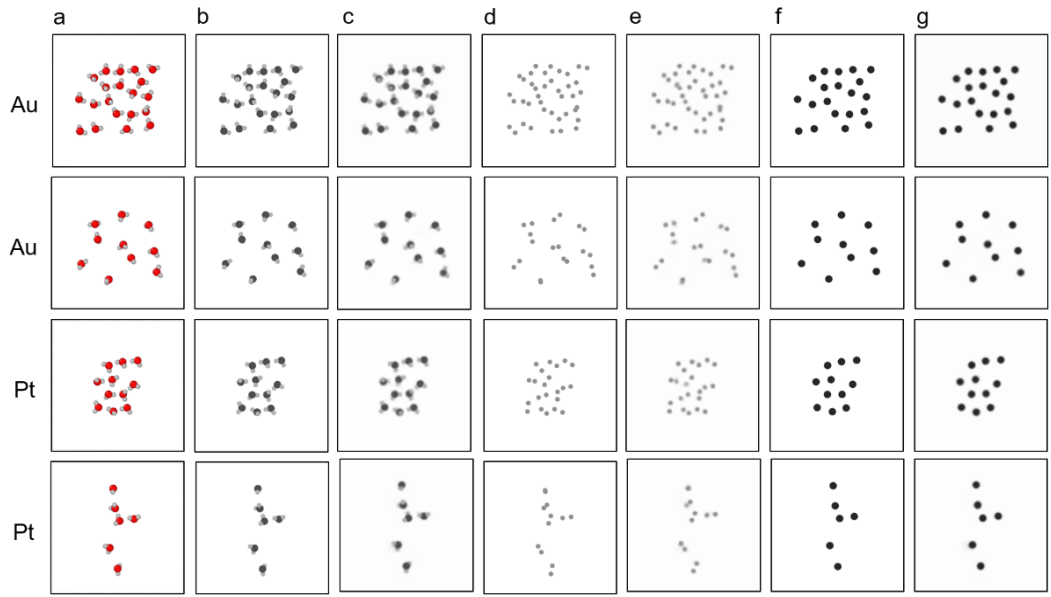

**Figure S4 | Detailed prediction and reference representations of Fig. 2. a,** The atomic model presented by OVITO. **b, d, f,** The reference representations of structure **a**, which is atom type, positive charge and negative charge, respectively. **c, e, g,** The representations predicted by the NN for structure **a**, which is also atom type, positive charge and negative charge, respectively.

## 9 NN Prediction of hydration Sodium

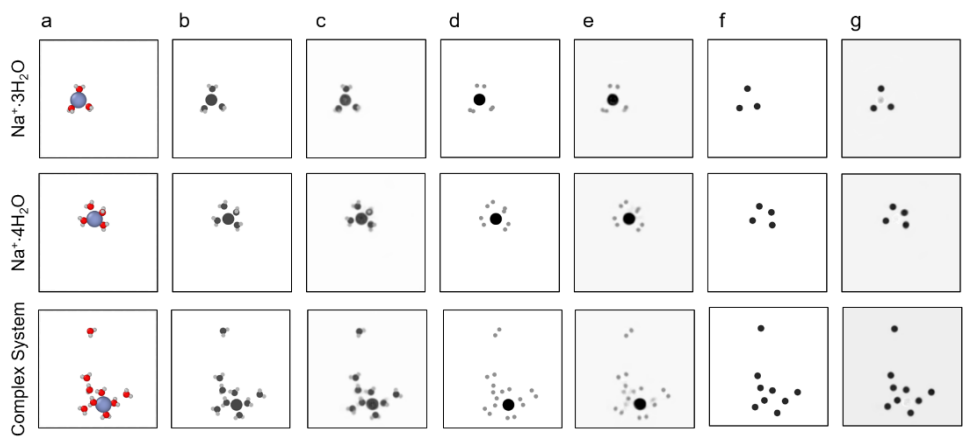

**Figure S5 | Detailed prediction and reference representations of Fig. 3. a,** The atomic model presented by OVITO. **b, d, f,** The reference representations of structure **a**, which is atom type, positive charge and negative charge, respectively. **c, e, g,** The representations predicted by the NN for structure **a**, which is also atom type, positive charge and negative charge, respectively.

## 10 Bad prediction of NN on interfacial water (Au surface) data

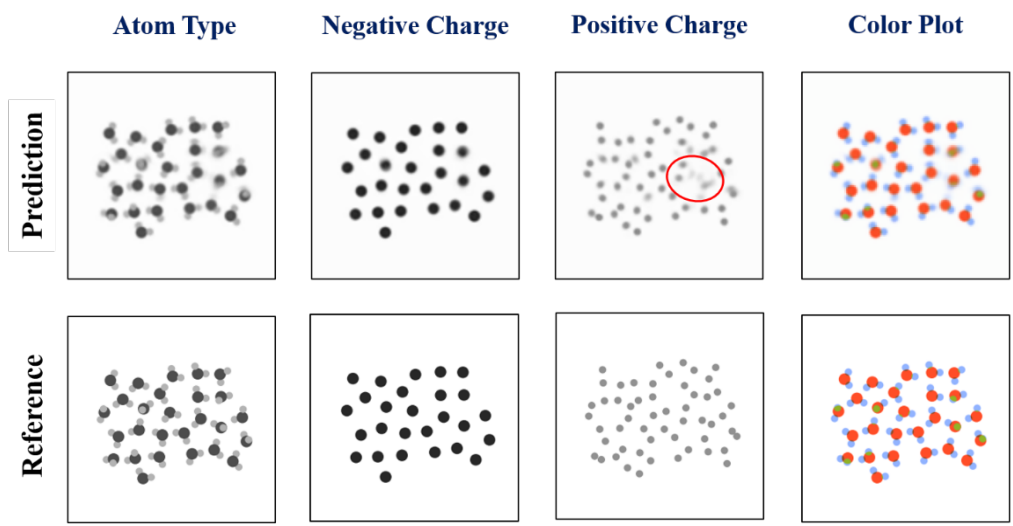

**Figure S6 | Bad prediction of NN on interfacial water (Au surface) data.** The representations of structure **a**, which is atom type, negative charge, positive charge, and color plot, respectively (the upper panels, predicted by NN; the lower panels, reference representations).

## 11 Correction of the hydrogen atom shift by transfer learning

After training the NN using data of interfacial water on Au (111) substrate, we test the transferability of NN using the data of interfacial water on Pt (111) substrate (testing data). Manually examination of the prediction of NN on testing data revealed that some of the predicted hydrogen atom positions drifted along hydrogen bonds compared to the reference. As can be seen from fig. S7 (the middle plot), the hydrogen shifts are marked with arrows, and the misdescription of hydrogen atom is marked with the circle. After re-training (transfer learning) the NN using 5000 data of interfacial water on Pt (111) substrate, the hydrogen shifts almost disappeared (fig. S7, the right plot).

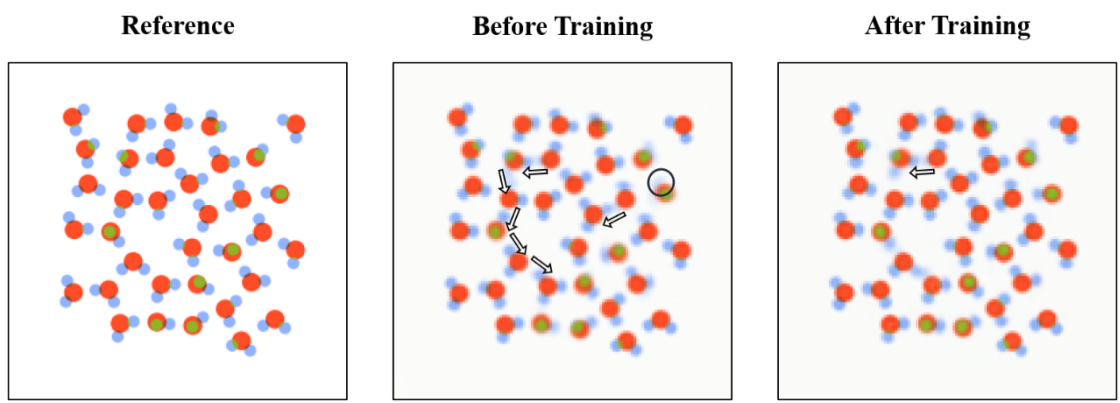

**Figure S7 | Hydrogen atom shift in NN prediction of the water structure on Pt surface.** The first plot is the reference structure; the second and third plots are the colormaps predicted by the NN before and after training on the interfacial water structures on Pt surface. The arrows indicate the directions of the displacement of hydrogen atoms, and the circles indicate the mispredictions of the hydrogen atom compared to the reference.

## 12 Prediction for experimental AFM images of $Na^+ \cdot 4H_2O$

Before prediction, the experimental AFM images to be predicted should be pre-processed. They should be padded or cut to the same size as the detection space of the training data, so as to ensure that the size of the imaging of a specific molecule relative to the detection space remains unchanged. Here, the detection space of experimental AFM images to be predicted is $1.8\ nm\ \times 1.8\ nm$ and the image size is $240 \times 240$ pixels, which should be padded to $333 \times 333$ pixels, since the detection space of the training data is $2.5\ nm\ \times 2.5\ nm$. Then, the images are resized to $128 \times 128$ pixels as the input data. In addition, to avoid the influence of experimental noise on the NN prediction, we adjust the noise amplitude of the data augmentation $c = 0.01$ during the NN training process for ionic hydrates data.

According to the prediction of NN of experimental data (Fig. 3, the prediction in last row), we create a $Na^+ \cdot 4H_2O$ structure on NaCl substrate and simulate the AFM images. Comparing the simulated (fig. S8, second row) and experimental (fig. S8, first row) AFM images, we can conclude that the NN can give a good prediction for Na hydrates experimental data. To be noticed, the unsymmetric feature of the experimental AFM image was caused by the tip drift during the image scanning. Specifically, it took a long time (about 1 hour) to complete the scanning of one AFM image. In such a long time, the tip would drift to a certain extent, and the AFM image would appear unsymmetric.

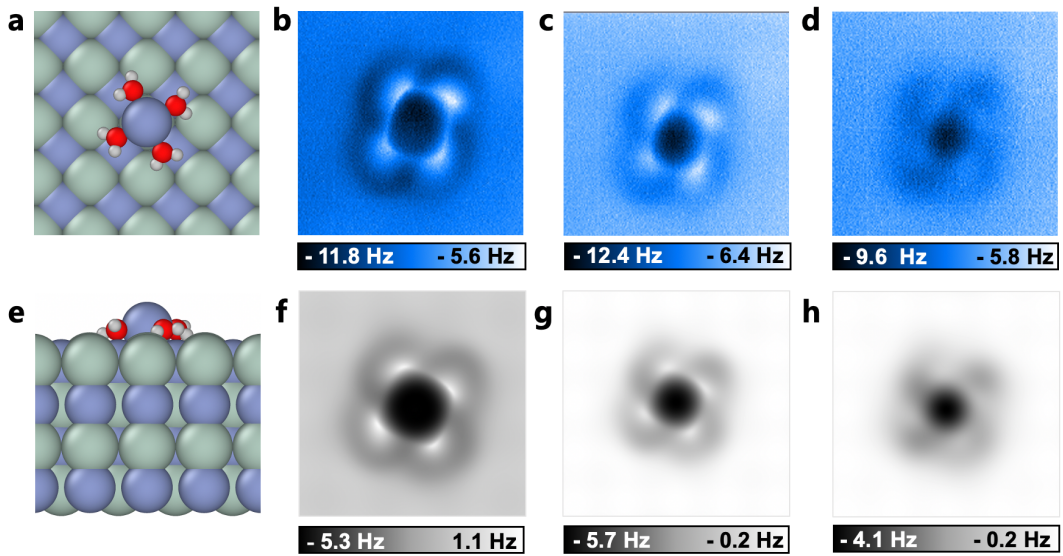

**Figure S8 | Validation of the predicted Na$^+$·4H$_2$O structure according the experimental AFM images. a** and **e,** Atomic model (a: top view; e: side view) according to the NN prediction in Fig. 3. The red, white, lilac and cyan spheres represent O, H, Na and Cl, respectively. **b - d,** Experimental AFM images of the Na$^+$·4H$_2$O at different tip heights of 25 pm (b), 70 pm (c) and 120 pm (d). **f - h,** Simulated AFM images of the Na$^+$·4H$_2$O at increasing tip heights from left to right. The tip heights in b-d are referenced to the STM set point on the NaCl surface (100 mV, 50 pA). The tip heights in f-h are defined as the vertical distance between the apex atom of the metal tip and the outmost atom of NaCl substrate. The oscillation amplitude of experimental and simulated images is 100 pm.

## 13  Prediction for Na hydrates whose water molecules are at different positions

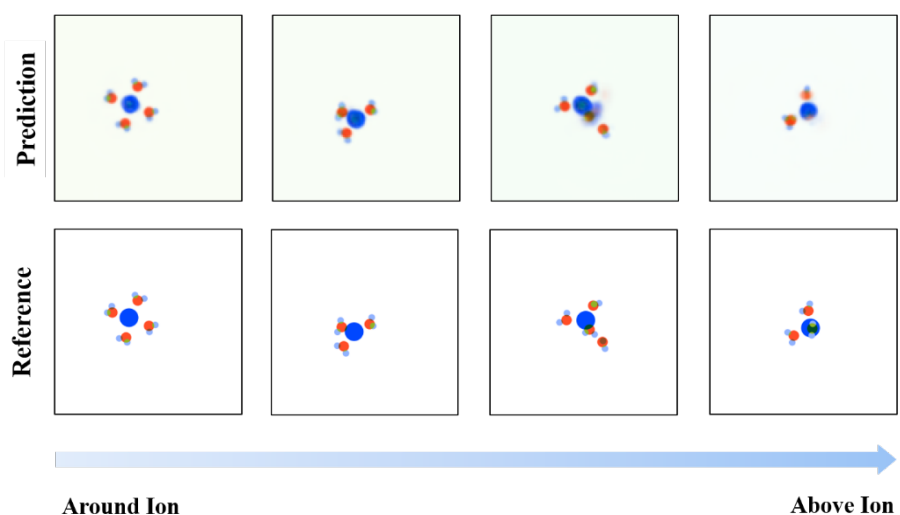

**Around Ion** → **Above Ion**

**Figure S9 | Prediction for Na hydrates whose water molecules are at different positions.** The representations of Na hydrates, from left to right where one of the water molecules posited from around the ion to above the ion (the upper panels, predicted by NN; the lower panels, reference representations).

## 14 Structure prediction of K$^+$ hydrates with Transfer Learning

To verify the general applicability of our transfer learning method, we carried out further investigations on K$^+$ hydrates, following the same operation as for Na$^+$ hydrates. In particular, thousands of K$^+$·nH$_2$O (n>3) data were prepared by simulation for NN transfer learning. To reproduce most of the important features of experimental AFM images, the AFM simulation parameter of charge at the tip apex was used Q = -0.20 e. After 50 epochs of training, the NN shows quite good performance on the K$^+$ hydrates testing data (fig. S10). The predicted accuracies for K, O and H (calculated from 700 simulated testing data) are 99.9%, 96.3%, and 81.4%, respectively. This is a good demonstration of the robustness and general applicability of our method.

Prediction of experimental AFM images of K$^+$ hydrates on Au surface is presented in the fourth row of fig. S10 below. Validation of the predicted K$^+$ hydrates structure is shown in fig. S11, which is verified by the consistency between the experimental AFM image and the simulated AFM image of the NN-predicted structure after DFT relaxation (fig. S11). It is worth mentioning that since the experimental image used to predict the structure is a part of a large interfacial hydrogen bond network, the simulated AFM images of the predicted structure near the boundary where the periodic boundary condition was applied, show some differences compared with the experimental image. Nevertheless, the AFM simulation of the predicted structure agrees well with the experimental result, demonstrating the high predictive power of our method for the experimental AFM results.

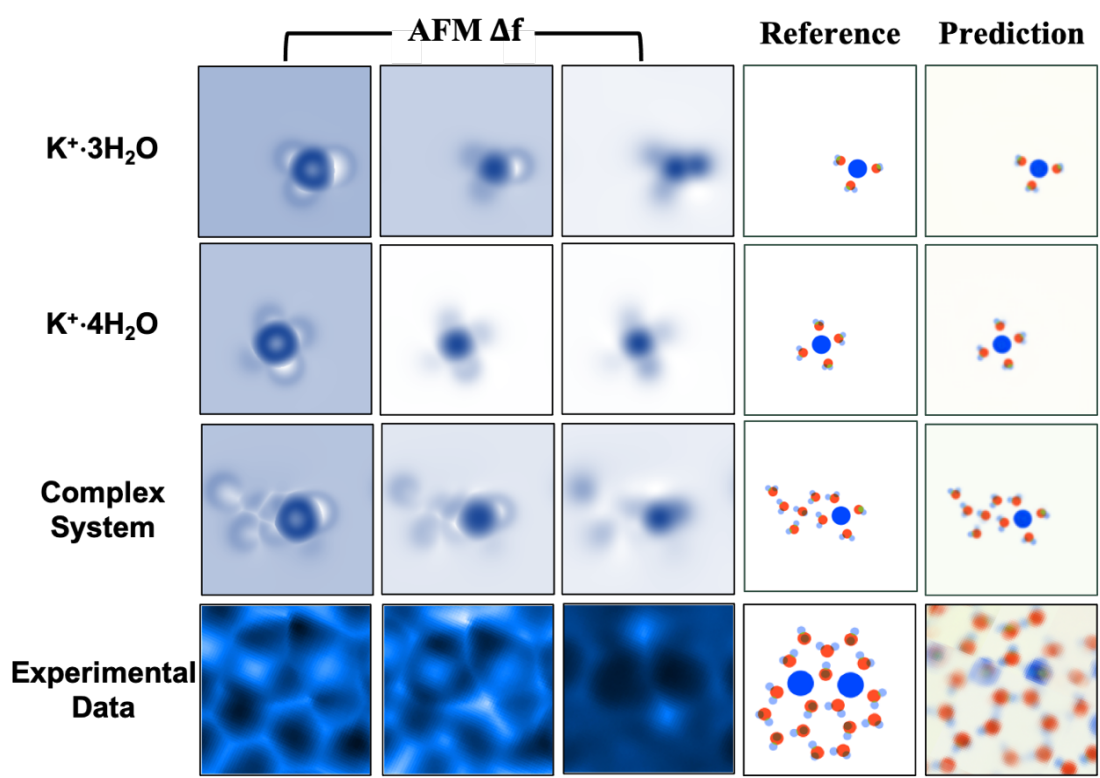

**Figure S10 | Examples of network prediction of simulated and experimental data for $K^+$ hydrates.** Rows 1-3 are the $K^+·3H_2O$, $K^+·4H_2O$, and the complex $K^+$ hydrate structure, selected from the calculated test dataset, respectively. Row 4 is the experimental data. Columns 1-3 are AFM images (input data) with increasing tip-sample distance. Column 4 is the reference a-vdW sphere representation of the structure (label), and column 5 is the prediction from the network. For the row of Experimental Data, the AFM images are obtained at different tip heights of -235 pm (left panel), -205 pm (middle panel) and -145 pm (right panel). The tip heights in experimental data are referenced to the STM set point on the $K^+$ cation (100 mV, 10 pA). The reference structure was verified by relaxing the predicted structure by DFT and comparing the AFM images simulated based on it with the experiment images. The red and indigo spots represent oxygen and kalium atoms, respectively. The light blue, green, and brown spots represent hydrogen atoms at the same height as, above and below the oxygen atoms, respectively.

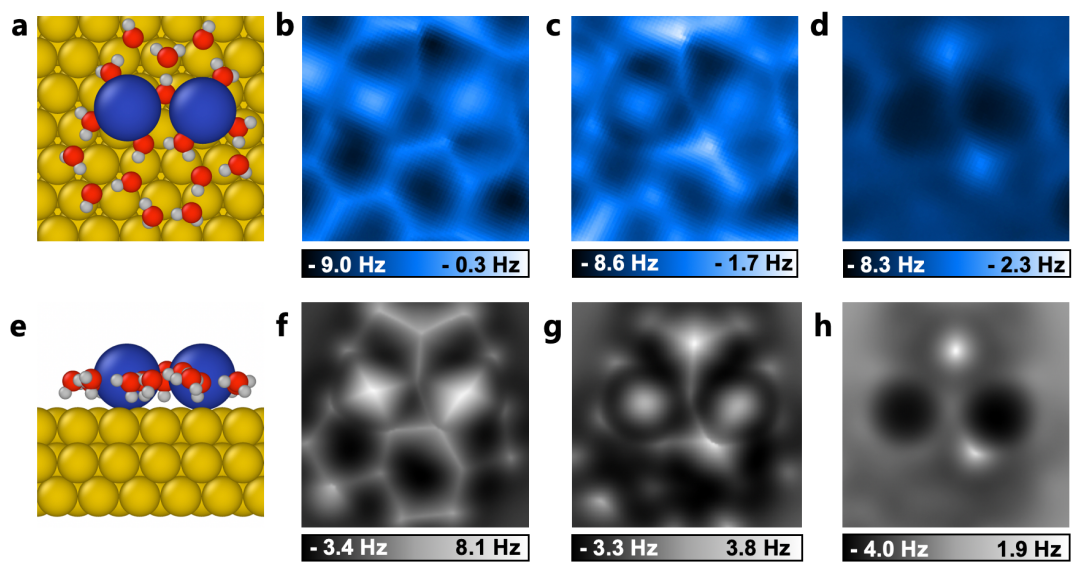

**Figure S11 | Validation of the predicted K$^+$ hydrates structure.** (a and e), Atomic model (a: top view; e: side view) according to the NN prediction of experimental data in fig. S10. The red, white, indigo and golden spheres represent O, H, K and Au, respectively. (b-d), Experimental AFM images of the K$^+$ hydrates at different tip heights of -235 pm (b), -205 pm (c) and -145 pm (d). (f-h), Simulated AFM images of the K$^+$ hydrates at increasing tip heights from left to right. The tip heights in b-d are referenced to the STM set point on the K$^+$ cation (100 mV, 10 pA). The tip heights in f-h are defined as the vertical distance between the apex atom of the metal tip and the outmost atom of Au substrate. The oscillation amplitude of experimental and simulated images is 100 pm.

## 15 The prediction performance of NN with or without transfer learning

We compare the prediction performance of NN with or without transfer learning as the training dataset size varies. The same validation dataset is used regardless of how much training data there is and whether or not the NN training uses transfer learning techniques. As shown in fig. S12, we chose the structures on which the NN without transfer learning performs best according to MSE, however, the NN trained without transfer learning still cannot predict the position of hydrogen atoms and the water molecule orientation.

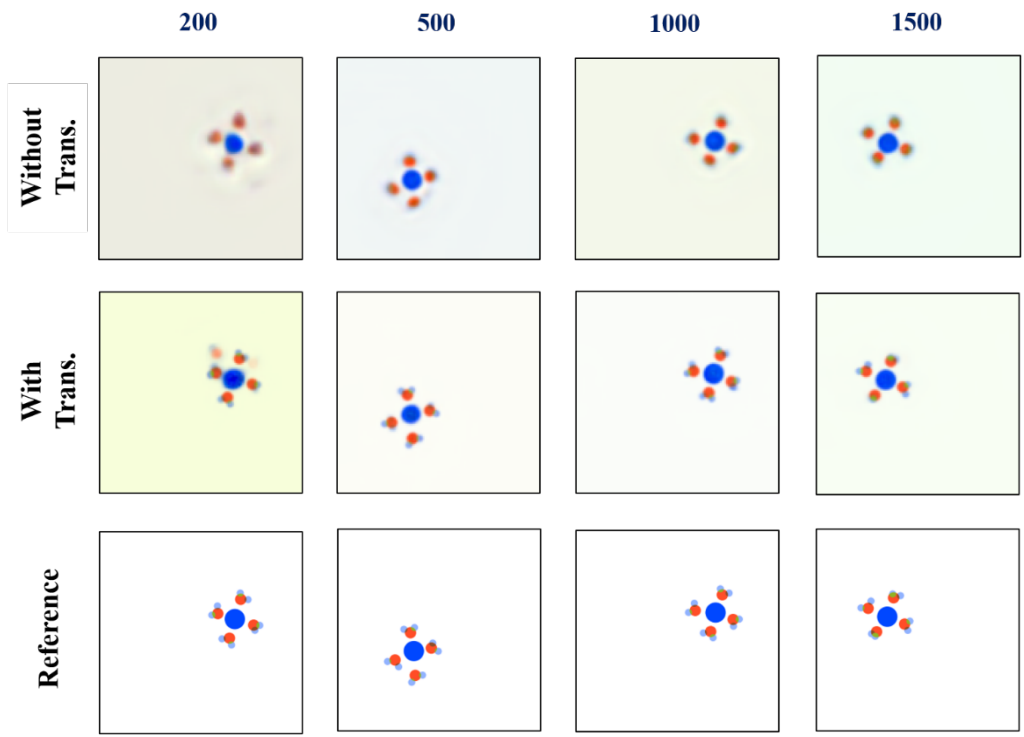

**Figure S12 | Prediction of Na hydrates by NN with or without transfer learning.** The first and second rows are the representation of the $Na^+ \cdot 4H_2O$ predicted by NN with or without transfer learning, respectively; the third row is the reference representation. The number of training data of NN varies from left to right and is denoted at the top of each column. For each column, the structure on which the NN without transfer learning performs best according to MSE is chosen.

## 16 Object detection for each atom

Negative and positive charge maps are used for atom detection. For the detection of oxygen atoms and sodium ions, we take pixel grayscale and pixel number as the criteria for atom detection and identification. For each element, any pixel with grayscale $g_p$ satisfying $g_p \in [g_{min}^i, g_{max}^i]$ can be considered as part of it. After finding all relevant pixels, we consider all connected pixels as one possible atom and count their number $N$. If the number of those pixels satisfies $N \in [N_{min}^i, N_{max}^i]$, this block of pixels can be regarded as an atom $j$ representing element $i$. To obtain the coordinates of each detected atom $j$, we average the coordinates of the pixels belonging to it and set the average to $r_j$. For each atom $j$, we calculate its distance from the corresponding atom in the label. Once the distance is less than the threshold $d_{match}$, we consider it as a true positive (TP) prediction. Missing and mismatched predicted atoms are treated as false negative (FN) and false positive (FP) prediction, respectively. Since the atoms are detected one by one, to avoid one detected atom affecting the recognition of other atoms, we use a circle mask of radius $r$ to cover the detected atoms.

For the hydrogen atoms, specifically, only a few pixels of light grayscale are presented. We use the edge detection for atom identification. The image is fist smoothed by operator $S_s$

$$S_s = \begin{pmatrix} 0 & 1 & 0 \\ 1 & 6 & 1 \\ 0 & 1 & 0 \end{pmatrix} \quad (S6)$$

Then, we use 2 sober operators, $S_x$ and $S_y$, to compute the gradients on the x-axis and y-axis, resulting in scalar gradients at each position. Where the gradients falling into the range $[gr_{min}, gr_{max}]$ can be considered as edges. The pixels of the region bounded by each connected edge should also be within the range $[N^H_{min}, N^H_{max}]$. Furthermore, we should consider the shape of the region as an important criterion. We define the shape factor of a region as follows:

$$f = \sum_i R_i^2 / (\sum_i R_i)^2, \quad (S7)$$

where $R_i$ denotes the distance from the i-th pixel to the center of the region, and the sum is over all pixel at the edge of the region. A perfect circle has a shape factor of 1. Once the $f > f_{max}$ where $f_{max}$ is a threshold, the shadow should be discarded. In addition, if two hydrogen atoms overlap in the label and we can only detect one region in the prediction, we count them as two TPs.

Finally, for each element, the prediction accuracy is calculated by eq. S7 based on all predictions for it in the target dataset.

$$accuracy = \frac{TP}{TP+FP+FN}. \quad (S8)$$

To avoid overestimating the performance of the NN, the parameter set in the prediction accuracy calculation algorithm was adjusted and cross-checked by manual calculated prediction accuracy of each atom species. We manually identified the atoms and calculate the prediction accuracy for each atom species among different datasets. Calculated using the same datasets used in Fig. 4, we plot the error bars with the manually collected data (see fig. S13). The prediction accuracy of machine calculation of this parameter set is slightly lower than that of manual calculation, indicating that the prediction accuracy of the machine calculation is reliable.

**Table S4 Parameters used in atom detection**

| Element | H | O | Na/K |
|---|---|---|---|
| $g_{min}$ |  | 25 | 0 |
| $g_{max}$ |  | 50 | 75 |
| $N_{min}$ | 9 | 3 | 10 |
| $N_{max}$ | 50 | 100 | 200 |
| $gr_{min}$ | 23 |  |  |
| $gr_{max}$ | 96 |  |  |
| $d_{match}$[Å] | 0.2 | 0.2 | 0.4 |
| $r$[Å] | 0.6 | 0.8 | 1.6 |
| $f_{max}$ | 1.15 |  |  |

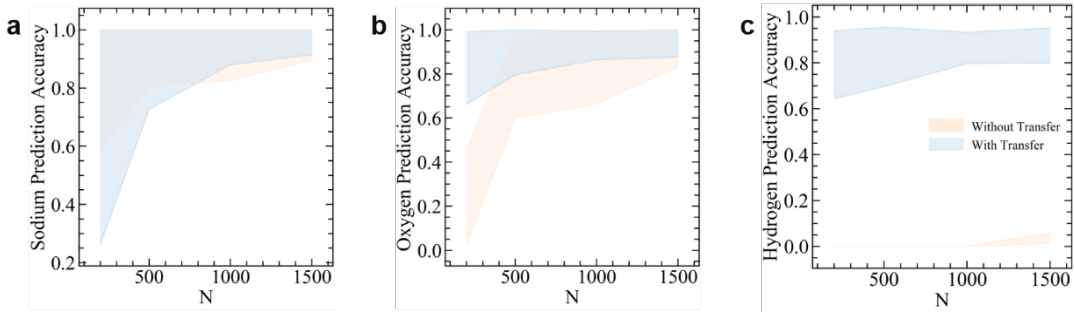

**Figure S13 | The prediction accuracy error bar in Fig. 4 with the manually collecting data. a**, **b** and **c** are the positional accuracies of sodium, oxygen and hydrogen atoms, respectively. The blue and orange color patched represent the training process with and without pretrained parameter loading, respectively. The upper and lower bounds of the color patches are manually calculated from the data at the top and bottom 20% of the loss ranking, respectively.

# Reference

1. Çiçek, Ö, Abdulkadir, A, Lienkamp, SS, *et al*. 3D U-Net: learning dense volumetric segmentation from sparse annotation. In: *International conference on medical image computing and computer-assisted intervention, 2016*, p. 424-32. Springer.
2. Litjens, G, Kooi, T, Bejnordi, BE, *et al*. A survey on deep learning in medical image analysis. *Medical image analysis*. 2017; **42**: 60-88.
3. Plimpton, S. Fast parallel algorithms for short-range molecular dynamics. *Journal of computational physics*. 1995; **117**(1): 1-19.
4. AMBER 14;University of California, San Francisco, 2014.
5. Fuentes-Azcatl, R, Barbosa, MC. Sodium chloride, NaCl/epsilon: new force field. *J Phys Chem B*. 2016; **120**(9): 2460-70.
6. Leontyev, IV, Stuchebrukhov, AA. Polarizable molecular interactions in condensed phase and their equivalent nonpolarizable models. *J Chem Phys*. 2014; **141**(1): 014103.
7. Yagasaki, T, Matsumoto, M, Tanaka, H. Lennard-Jones parameters determined to reproduce the solubility of NaCl and KCl in SPC/E, TIP3P, and TIP4P/2005 water. *Journal of Chemical Theory and Computation*. 2020; **16**(4): 2460-73.
8. Peng, JB, Cao, DY, He, ZL, *et al*. The effect of hydration number on the interfacial transport of sodium ions. *Nature*. 2018; **557**(7707): 701-5.
9. Ryckaert, JP, Ciccotti, G, Berendsen, HJC. Numerical integration of the cartesian equations of motion of a system with constraints: molecular dynamics of n-alkanes. *Journal Of Computational Physics*. 1977; **23**(3): 327-41.
10. Perdew, JP, Burke, K, Ernzerhof, M. Generalized gradient approximation made simple. *Phys Rev Lett*. 1996; **77**(18): 3865.
11. Hamann, D. H 2 O hydrogen bonding in density-functional theory. *Phys Rev B*. 1997; **55**(16): R10157.
12. Kresse, G, Furthmüller, J. Efficient iterative schemes for ab initio total-energy calculations using a plane-wave basis set. *Phys Rev B*. 1996; **54**(16): 11169.
13. Klimeš, J, Bowler, DR, Michaelides, A. Chemical accuracy for the van der Waals density functional. *Journal of Physics: Condensed Matter*. 2009; **22**(2): 022201.
14. Klimeš, J, Bowler, DR, Michaelides, A. Van der Waals density functionals applied to solids. *Phys Rev B*. 2011; **83**(19): 195131.
15. Jorgensen, WL, Tirado-Rives, J. The OPLS [optimized potentials for liquid simulations] potential functions for proteins, energy minimizations for crystals of cyclic peptides and crambin. *Journal of the American Chemical Society*. 1988; **110**(6): 1657-66.
16. Stukowski, A. Visualization and analysis of atomistic simulation data with OVITO–the Open Visualization Tool. *Modelling and simulation in materials science and engineering*. 2009; **18**(1): 015012.
17. Kingma, DP, Ba, J. Adam: A method for stochastic optimization. *arXiv preprint arXiv:14126980*. 2014.